# TransCDR: a deep learning model for enhancing the generalizability of cancer drug response prediction through transfer learning and multimodal data fusion for drug representation

Xiaoqiong Xia[a], Chaoyu Zhu[b], Yuqi Shan[d], Fan Zhong[b,*], and Lei Liu[b,c,*]

[a] Institutes of Biomedical Sciences, Shanghai Medical College, Fudan University, Shanghai 200032, China
[b] Intelligent Medicine Institute, Shanghai Medical College, Fudan University, Shanghai 200032, China
[c] Shanghai Institute of Stem Cell Research and Clinical Translation, Shanghai, 200120, China
[d] Shanghai World Foreign Language Academy, Shanghai, 200233, China

*Corresponding authors
E-mail addresses: zonefan@163.com (F. Zhong) and liulei_sibs@163.com (L. Liu)

## Abstract

Accurate and robust drug response prediction is of utmost importance in precision medicine. Although many models have been developed to utilize the representations of drugs and cancer cell lines for predicting cancer drug responses (CDR), their performances can be improved by addressing issues such as insufficient data modality, suboptimal fusion algorithms, and poor generalizability for novel drugs or cell lines. We introduce TransCDR, which uses transfer learning to learn drug representations and fuses multi-modality features of drugs and cell lines by a self-attention mechanism, to predict the $IC_{50}$ values or sensitive states of drugs on cell lines. We are the first to systematically evaluate the generalization of the CDR prediction model to novel (i.e., never-before-seen) compound scaffolds and cell line clusters. TransCDR shows better generalizability than 8 state-of-the-art models. TransCDR outperforms its 5 variants that train drug encoders (i.e., RNN and AttentiveFP) from scratch under various scenarios. The most critical contributors among multiple drug notations and omics profiles are Extended Connectivity Fingerprint and genetic mutation. Additionally, the attention-based fusion module further enhances the predictive performance of TransCDR. TransCDR, trained on the GDSC dataset, demonstrates strong predictive performance on the external testing set CCLE. It is also utilized to predict missing

CDRs on GDSC. Moreover, we investigate the biological mechanisms underlying drug response by classifying 7,675 patients from TCGA into drug-sensitive or drug-resistant groups, followed by a Gene Set Enrichment Analysis. TransCDR emerges as a potent tool with significant potential in drug response prediction. The source code and data can be accessed at https://github.com/XiaoqiongXia/TransCDR.

## Keywords

# 1 Introduction

Tumors exhibit intra- and inter-tumoral heterogeneity [1], contributing to the variable efficacy of anticancer drugs among different tumor subtypes and patients. In order to enhance clinical outcomes and patient survival rates, precision/personalized medicine [2] seeks to individualize treatments based on the specific molecular characteristics of each patient [3]. Genomics, epigenomics and transcriptomics have emerged as invaluable tools for providing unprecedented insights into the underlying molecular mechanisms of cancer [4]. Precision medicine and drug repurposing can be considerably facilitated by performing a systematic analysis of drug properties and multi-omics features of cancer cell lines and accurately predicting cancer cell drug responses.

The advent of large-scale drug sensitivity data and the genomic data for over 1,000 cultured cancer cell lines, such as Drug Sensitivity in Cancer (GDSC) [5], NCI-60 [6] and Cancer Cell Line Encyclopedia (CCLE) [7], has enabled the development of computational models to predict cancer drug responses (CDR). Several novel models, including DeepCDR [8], DeepTTA [9] and GraphDRP [10], have been reported using standard datasets extracted from the GDSC. These end-to-end models share a similar architecture, with drug and cell line encoders learning representations for drugs and cell lines. Stacked fully connected layers then utilize these representations to predict drug sensitivities. Consequently, generating an accurate and robust prediction model requires appropriate representation learning for drugs and cell lines. The emergence of novel deep learning modules, such as convolutional neural networks (CNN), graph neural networks (GNN) [11, 12], and Transformer [13], has motivated their application to CDR models [14]. For example, GraphDRP [10] and GraOmicDRP [12] utilized GNNs (e.g., GIN and GAT) to learn drug features from their graph representation. DeepTTA employed a Transformer module for drug representation learning from SMILES strings [9]. CNN blocks were utilized to extract features from multi-omics data for cell lines [10, 12, 15]. A thorough comparison of the state-of-the-art (SOTA) models demonstrated that GraphDRP and DeepCDR outperformed traditional machine learning methods (e.g., ENet and random forest) and 3 deep learning methods (i.e., CDRscan, tCNN and MOLI) [16]. Additionally, models such as TGSA [17] and DRPreter [18] were proposed to make better use of prior domain knowledge (e.g., protein–protein interaction). They applied GNN to extract cell line features from gene networks.

Despite the considerable progress achieved in CDR models, several limitations still exist. Firstly, labeled drugs for CDR tasks are often scarce, leading to deficient representation learning of drugs. Secondly, while these CDR models aim to learn more appropriate drug representations from 1D Simplified Molecular Input Line Entry System (SMILES) strings [19] or 2D molecular graphs, or extended-connectivity fingerprints (ECFPs) and achieve high accuracy, the potential interplay among multiple drug representations has yet to be fully explored [20]. Thirdly, the fusion representation of CDR is obtained by concatenating representations of drugs and cell lines, thereby limiting CDR models' performance. Finally, the accuracy of prior methods significantly drops when predicting the response of an unrepresented drug in the training set, and their inability to accurately predict CDRs in cold start scenarios has not been thoroughly evaluated and discussed. These substantial limitations will hinder the effectiveness of CDR models in precision medicine and drug repurposing.

Transfer learning is a technique that aims to enhance models' performance on small-volume datasets by transferring knowledge extracted from related large-scale datasets [21]. Although this technique is wildly applied in natural language processing [22] and computer vision, its development in computational chemistry is yet to be effectively realized. Recently, several pre-trained drug encoders have been made available. For example, ChemBERTa is a BERT-like transformer model pre-trained on a vast corpus of SMILES strings through masking language modeling of chemical SMILES strings [23]. Gin_supervised_masking is a graph isomorphism network (GIN) model pre-trained with supervised learning and attribute masking [24]. These pre-trained drug encoders can be implemented to learn global and expressive drug representation and transferred to various downstream tasks, such as drug response prediction, drug-target prediction, drug design, and property prediction.

We proposed an end-to-end regression/classification model, TransCDR (Figure 1), to overcome the abovementioned limitations. TransCDR captured high-dimensional features from the drug's SMILES strings (S), molecular graphs (G), and ECFPs (FP), as well as the associations between drug and cell line representations, to predict the $IC_{50}$ value when presented with a drug-cell line pair. TransCDR significantly outperformed SOTA models for predicting $IC_{50}$ values or sensitive states under warm and cold start. Several innovative aspects of the model's architecture contributed to the success of TransCDR. First, we introduced transfer learning to extract the chemical features of drugs automatically. Second, we integrated 3 drug structural representations (i.e., S, G, FP). Third, we leveraged a multi-head attention mechanism to fuse the representations of drugs and cell lines. Finally, we evaluated the prediction ability of TransCDR on external verification sets: CCLE and applied the trained TransCDR to screening drugs for clinical patients. Furthermore, we elucidated the biological mechanisms of candidate CDRs via Gene Set Enrichment Analysis (GSEA). Thus, TransCDR contributed to cancer drug prediction and drug repurposing/discovery.

## 2 Methods and materials

### 2.1 Data preparation

This study utilized GDSC, CCLE, and TCGA datasets. Specifically, GDSC was employed to assess the effectiveness of TransCDR across various application scenarios, including predicting missing CDRs for known cell lines and drugs and unseen cell lines, drugs, and both unseen cell-drug combinations. Additionally, all CDRs from the GDSC database were utilized for training the final TransCDR model, which was subsequently evaluated on external datasets: CCLE.

GDSC v2 constitutes a vital asset in the endeavor of discovering therapeutic biomarker for cancer cells [5]. We have gathered a total of 156,813 CDRs that satisfied three specific criteria, including 851 cancer cell lines and 225 drugs. 1) CDRs encompassed drug sensitivity profiles ascertained through the measurement of the half maximal inhibitory concentration ($IC_{50}$) or sensitive state, which indicated the capacity of a drug to impede the growth of specific cell lines. 2) The selected CDRs exhibited the presence of three omics data sets: genetic mutation, gene expression, and DNA methylation for the corresponding cell lines. 3) The included drugs possessed SMILES strings.

This study obtained mutation and copy number aberration (MC), gene expression (GE), and DNA methylation (DM) profiles from GDSC. Specifically, gene expression profiles were downloaded for 1,000 human cancer cell lines using transcriptional profiling arrays E-MTAB-3610, and were pre-processed using the R package *affy*. The Affymetrix GeneChip system, along with the robust multiarray average method, was employed for measuring gene expression [25], resulting in 18,451 gene expression values for each cell line. Subsequently, the gene expression matrix was then normalized using *z*-score. The MC data consisted of a binary matrix with 735 features, where 1 indicated a mutation or copy number aberration in the gene, and 0 indicated absence of such aberrations. The DM matrix was obtained by downloading the processed matrix of GSE68379 from GEO, where continuous values represented the methylation score of each CpG. The methylation scores of CpG sites were then averaged to obtain methylation scores for genes, resulting in 20,617 methylation values for each cell line. The DM matrix was also normalized by *z*-score. Drug SMILES strings were retrieved from PubChem [26] and converted to canonical SMILES using open-source cheminformatics software RDKit.

For each combination of cell line and drug $CDP_{ij}$, cell line $i$ was characterized using 3 types of omics data (i.e., MC, GE, DM); drug $j$ was represented by SMILES strings, and the label of $CDP_{ij}$ was the natural logarithm-transformed $IC_{50}$. A total of 156,813 CDPs were utilized in the development of the regression model. In classification experiments, $IC_{50}$ values were binarized based on the provided threshold for each drug [27]. Consequently, a total of 154,603 CDPs were obtained, with $CDP_{ij} \in \{0,1\}$, consisting of 18,143 sensitive CDPs and 136,460 resistant samples.

For the CCLE dataset, this study accessed MC, GE, and DM profiles as well as pharmacological profiling files from the Broad DepMap Portal. The processing steps for CDPs in GDSC were followed to extract 9,242 CDPs, which consisted of 401 cancer cell lines and 24 drugs, with $IC_{50}$ values transformed via natural logarithm. For the TCGA dataset, a total of 7,675 patients with multi-omics profiles, including MC (MC3 gene-level non-silent mutation), GE (Illumina HiSeq), and DM (Methylation 450k) were obtained from UCSC Cancer Genome Browser Xena [28] using TCGA patient ID. Due to differences in the feature dimensions of MC, GE, and DM between CCLE, TCGA and GDSC, the features of CCLE and TCGA were aligned with those of GDSC. Standardization of GE and DM profiled across different platforms was ensured through *z*-score normalization.

## 2.2 Data segmentation strategies

We employed 10-fold cross-validation (10-CV) to evaluate TransCDR's generalizability comprehensively. Datasets were divided based on 5 strategies: warm start, cold drug, cold scaffold, cold cell, and cold cell & scaffold.

1. Warm start: A warm start approach was adopted to assign a random selection of 80%, 10%, and 10% of the CDRs to the training, validation and testing sets, respectively. Notably, it was possible for a drug/cell line from the test or validation set to also be present in the training set. The models trained using the warm start strategy were then employed to predict the missing $IC_{50}$ values in the GDSC dataset.
2. Cold drug: Drugs present in the test/validation set were carefully excluded from the training set. Among the drug-associated CDRs, a random selection of 80% (180) drugs were assigned to the training set, 10% (22) to the validation set, and the remaining CDRs with 10% (23) drugs were designated for the test set. This experimental design aimed to assess the model's performance on unforeseen drugs. It was important to note that despite these efforts, there may be instances where different drugs share similar scaffolds, resulting in scaffold overlap between train, validation, and test data. Consequently, this overlap may potentially overestimate the generalization ability of the CDR model to novel drugs.
3. Cold scaffold: Initially, the Murcko scaffold of each SMILES was obtained using RDKit. Following this, the SMILES strings were clustered utilizing scaffold similarity. It was important to note that the Murcko scaffolds employed in the test/validation set were omitted from the training set. Consequently, a random allocation of 80% (170) scaffold-associated CDRs was allocated to the training set, with 10% (21) assigned to the validation set. The remaining CDRs, comprising 10% (19) drugs, were assigned to the test set. The purpose of this experiment was to assess the model's performance on unseen drugs with varying scaffolds.
4. Cold cell: Firstly, the cell lines were clustered based on 3 omics features using the $K$-means algorithm. The number of clusters used for comparisons in subsequent experiments was set at 10, 50, 100, and 200. Within this process, cell clusters allocated to the test/validation set were exclusively excluded from the training set. The training set consisted of a random selection of 80% cell line cluster-associated CDRs, while 10% were allocated to the validation set. The remaining CDRs, involving 10% of the cell line clusters, were assigned to the test set. This experiment aimed to evaluate the model's performance on previously unseen cell lines.
5. Cold cell & scaffold: The training set excluded both cell line clusters and drug scaffolds assigned to the test/validation set. For the training set, a random selection of 80% cell line cluster-associated and 80% drug scaffold-associated CDRs was allocated, whereas 10% cell line cluster-associated and 10% drug scaffold-associated CDRs were assigned to the test/validation set. This experiment assessed the model's performance on unseen cell line clusters and drug scaffolds.

## 2.3 Overall architecture of TransCDR

We proposed TransCDR, an end-to-end deep learning model which employed drugs' chemical structures and cell lines' multi-omics data to predict drug responses. TransCDR consisted of two prediction modes: regression for predicting and classification for predicting drug sensitivity or resistance on cell lines. The model was composed of four main components (Figure 1): (1) We employed ChemBERTa, a pre-trained model, to learn drugs' representations from SMILES strings, gin_supervised_masking, another pre-trained model, to learn drugs' molecular graph representations, and a stacked full connected (FC) layers module to acquire high-dimensional features from ECFPs. (2) We used three FCs to learn numerical representations of MC, GE, and DM data. (3) These drug and cell line representations were fused in a fusion module, a stacked multi-head attention layer module with 6 layers and 8 heads. The fusion module integrated multi-modality features of drugs and cell lines. (4) A regression/classification network with four FCs used fusion representations to predict drug responses. The components above of TransCDR were further elaborated in the subsequent paragraphs.

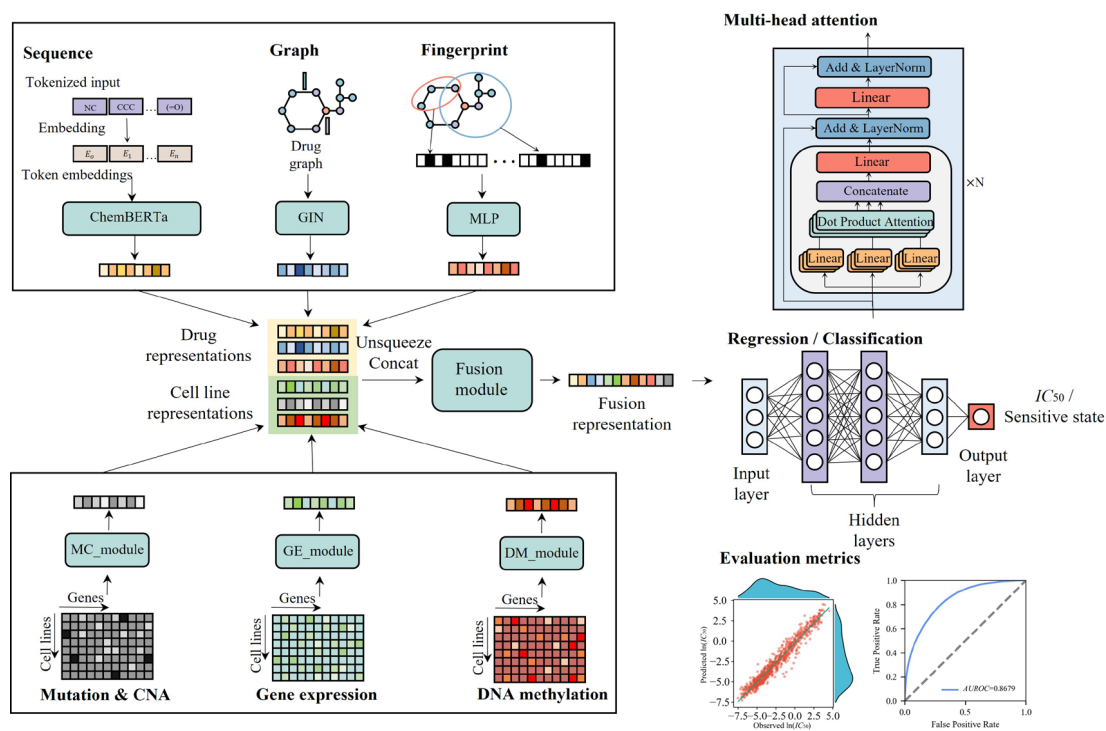

**Figure 1.** The framework of TransCDR includes three drug modules (ChemBERTa, GIN, MLP) for extracting drug features from SMILES strings, molecular graphs, and fingerprints, respectively. Similarly, there are three cell line modules (MC_module et al.) for extracting cell line features from genomic mutation, gene expression, and DNA methylation data, respectively. The drug and cell line representations are fused using a self-attention-based fusion module. Finally, the fusion representation is fed to a regression/classification network consisting of four fully connected layers to predict the $\ln(IC_{50})$ or sensitive state.

## 2.3.1 Drug representations

We employed three drug encoders to acquire numerical representations from the three basic molecular notations (S, G, and FP), followed by applying three notation-specific networks that extracted 256-dimension features from the numerical representations.

### 2.3.1.1 Sequence representation

We utilized the SMILES format to represent drugs, which involved a series of characters indicating atom and bond symbols and a few grammar rules resembling natural language. To this end, we proposed employing a pre-trained BERT-like model called ChemBERTa to acquire the numerical representations from SMILES strings. ChemBERTa [29] is pre-trained on 10M SMILES strings from PubChem using the masked language modelling approach. Figure 1 displays the specifications of the sequence representation module used in our experiments. Subsequently, the SMILE string was tokenized into sub-word token strings using the Byte Pair Encoding tokenizer, and then converted into token IDs with a maximum sequence length of 512. Next, the token IDs were inputted into the pre-trained ChemBERTa to obtain the sequence representation. The numeric representation of a SMILES string is computed as follows:

$$T = \text{tokenizer}(\text{SMILES}) \tag{1}$$

$$h_s = \text{ChemBERTa}(T) \tag{2}$$

Where $T$ indicates the token IDs of a SMILES string $T = \{t_1, t_2, \ldots, t_{512}\}$, and $h_s$ represents the learned numeric representation of the SMILES, with a dimension of 768. The ChemBERTa and tokenizer were downloaded from HuggingFace [30]. Furthermore, we extracted features from the numeric representations utilizing a neural network, with two hidden layers comprising 1,024 and 256 neural units, respectively. Every layer is formulated according to the following equation:

$$h_s = \text{ReLU}(W_i h_s + b_i) \tag{3}$$

Where $W_i$ and $b_i$ represent learnable matrices. The output size of the network is set to 256 to facilitate fusion operation.

### 2.3.1.2 Graph representation

The recent emergence and success of GNNs have inspired their application to drug representations. Specifically, we represented drugs as molecular graphs as $G = (V, E)$, where $V$ denotes the atoms, and $E$ denotes the chemical bonds node. Each node $v \in V$ is associated with node features $h_v$ and each edge $(u, v) \in E$ is associated with edge features $e_{uv}$

$$h_v^{(l+1)} = \text{MLP}^{l+1}\left((1 + \epsilon^{l+1}) * h_v^l + \sum_{u \in N(v)} e_{uv} * h_u^l\right) \tag{4}$$

$$h_g = (\frac{1}{N}\sum_N h'_v)) \tag{5}$$

$$h'_v = \text{CONCAT}(h_v^0, h_v^1, \dots, h_v^l) \tag{6}$$

Where $v$ represents the target node, $u$ represents the neighboring node of $v$, and $e_{uv}$ denotes the weight assigned to the edge from $u$ to $v$. The model includes a learnable parameter $\epsilon$ and employs $h_v^l$, the node representation of layer $l$, and $h_G$, the graph representation. The pre-trained GIN model, gin_supervised_masking, performed well in learning local and global representations at the individual node and whole graph level. To learn the appropriate representations with a dimension of 300 from molecular graphs, we applied a neural network with 2 hidden layers to extract features from these representations.

### 2.3.1.3 Fingerprint representation

The topological fingerprints of drugs were captured using ECFP representations [31] based on Morgan's algorithm with RDKit. Specifically, each atom was assigned a unique integer identifier and updated to represent larger circular substructures with a radius of 2. The final substructures were hashed into a binary vector with a length of 1,024, defined as $FP = \{fp_1, fp_2, \dots, fp_{1024}\}$, where $fp_i \in \{0,1\}$. ECFP features were also extracted using a neural network with 2 hidden layers.

### 2.3.1.4 TransCDR variants without pre-training

To examine the effectiveness of transfer learning, we replaced the pre-trained drug representation modules, namely ChemBERTa and gin_supervised_masking, with non-pre-trained modules, including CNN, RNN, AttentiveFP, NeuralFP, and ECFP. All parameters were initialized randomly and subsequently learned from scratch through a back-propagation algorithm. One-hot encoding was used to represent drugs in CNN and RNN, whereas AttentiveFP and NeuralFP represented drugs as pre-defined graph structures with atomic and bond features. The drug module architecture was identical to that of DeepPurpose [32].

### 2.3.2 Cell line representations

We employed a late fusion strategy to process high-dimensional and heterogeneous omics data and capture complex relationships from mutation and copy number aberration, gene expression, and DNA methylation profiles. We used omics-specific networks to extract features of cell lines from each omics and fuse these 3 types of features using multi-head attention. The fully-connected networks had 2 hidden layers with 1,024 and 256 neural units. We mapped the 3 types of omics data into a latent space with an embedded dimension fixed at 256.

$$h_{\text{MC}} = \text{Network}_{\text{MC}}(X_{\text{MC}}) \tag{7}$$

$$h_{GE} = \text{Network}_{GE}(X_{GE}) \tag{8}$$
$$h_{DM} = \text{Network}_{DM}(X_{DM}) \tag{9}$$

Where $h_{GE}$, $h_{MC}$, and $h_{DM} \in \mathbb{R}^{n*d}$, $d$=256, and $n$ is the batch size.

## 2.3.3 Multi-head attention for feature fusion

We proposed utilizing the multi-head attention mechanism to model the relationships between drug features (i.e., sequences, graphs and ECFPs) and cell line features (i.e., MC, GE, and DM). Initially introduced in Transformer [13], the multi-head attention method has been widely adopted for multi-modality fusion [33, 34]. Specifically, the attention module mapped a query and a set of key-value pairs to an output generated as a weighted sum of the values. The attention is formulated as follows:

$$\text{Attention}(Q, K, V) = \text{softmax}(\frac{QK^T}{\sqrt{d_k}})V \tag{10}$$

Where $Q, K, V \in \mathbb{R}^{n*6*d_k}$, derived from the concatenation of 3 drug and 3 cell line features, $n$ is the batch size, $d_k$ represents the feature dimension, T is a transpose operation. To learn the features from distinct representation subspaces, we projected the $Q$, $K$, and $V$ $h$ times, and calculated the multi-head attention function as follows:

$$Q_i = QW_i^Q + b_i^Q, i \in \{1,2,\ldots,h\} \tag{11}$$
$$K_i = KW_i^K + b_i^K, i \in \{1,2,\ldots,h\} \tag{12}$$
$$V_i = VW_i^V + b_i^V, i \in \{1,2,\ldots,h\} \tag{13}$$
$$\text{MultiHeadAtt}(Q, K, V) = \text{CONCAT}(head_1, head_2, \ldots, head_n)W^O \tag{13}$$
$$head_i = \text{Attention}(Q_i, K_i, V_i) \tag{14}$$

where $W_i^Q, W_i^K, W_i^V \in \mathbb{R}^{d*d_k}$, $W^O \in \mathbb{R}^{h*d_k*d}$ $b_i^Q, b_i^K$ and $b_i^V$ are learnable matrices, $h$ is set at 8, $d_k = \frac{d}{h} = \frac{256}{8} = 32$.

The multi-head attention mechanism was the primary constituent in constructing the fusion module of TransCDR. More specifically, the fusion module consisted of 6 identical multi-head attention layers. The output of this module was then flattened and incorporated into a regression module. Our study delved into the inquiry of 3 attention modules, namely self-attention, drug-cell line attention (DCA), and cell line-drug attention (CDA) (Figure 2).

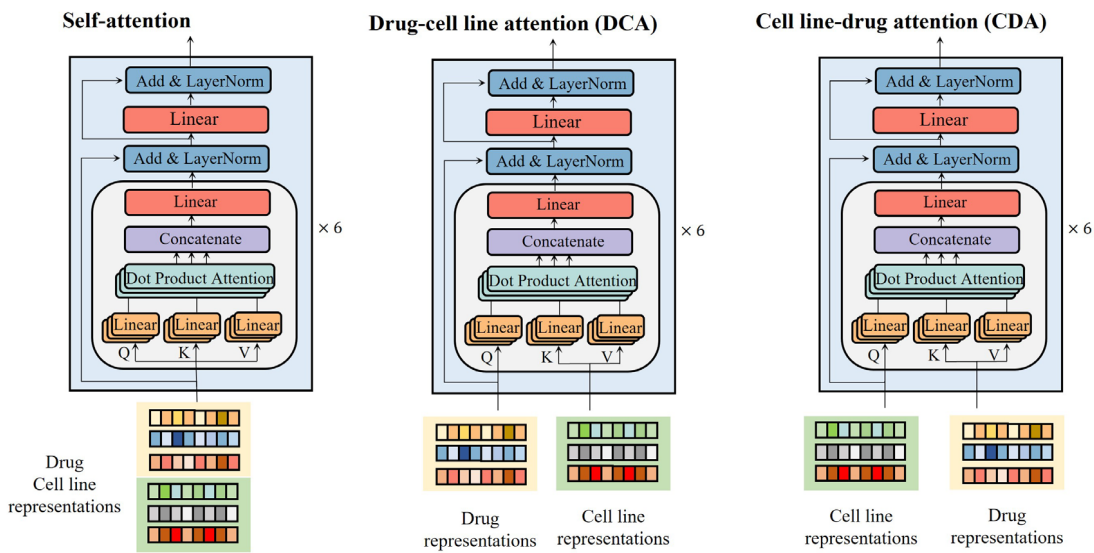

**Figure 2.** Illustration of the three attention modules employed in this study. (A) Self-attention: a module concatenates drug and cell line representations to serve as the Q, K, and V parameters. (B) Drug-cell line attention: Q denotes the drug representations, and K and V correspond to the cell line representations. (C) Cell line-drug attention: Q denotes the cell line representations, and K and V correspond to the drug representations.

### 2.3.4 Prediction module

The prediction module comprised a four-layer neural network incorporating rectified linear unit (ReLU) activation functions and dropout layers characterized by a dropout rate of 0.1. The output value pertained to a cell line's predicted drug response/sensitive state. To train regression/classification models, we adopted either the mean squared error (MSE) or binary cross entropy (BCE) loss function, which we subsequently back-propagated to the network and then updated all parameters end-to-end.

### 2.4 Performance metrics

For the regression experiments predicting $\ln(IC_{50})$ values of drugs and cell lines, we assessed TransCDR's performance using 4 evaluation measures: root mean square error (*RMSE*), Pearson correlation coefficient (*PC*), Spearman's rank correlation coefficient (*SC*), and concordance index (*C-index*). *RMSE* was used to calculate the difference between predicted and ground truth $IC_{50}$ values:

$$RMSE = \sqrt{\frac{1}{N}\sum(y_i - \widetilde{y_i})^2} \tag{15}$$

Where $N$ denotes the size of the test set. $y_i$ and $\widetilde{y_i}$ represent the ground truth and predicted $IC_{50}$ values, respectively. *PC* and *SC* measured the linear and rank-based correlations between ground truth and predicted $IC_{50}$ values. Additionally, we evaluated TransCDR's predictions using the *C-index*.

For the classification experiments, we evaluated the performance of each method using the Area under the Receiver Operating Characteristics (*AUROC*) and the Area

Under the Precision-Recall (*AUPR*) curves. *AUPR* was used as the primary metric, especially when negative samples were much more extensive than positive ones [35].

Lastly, we employed the two-sided Wilcoxon rank sum test with a significance threshold 0.05 to demonstrate the significant performance difference between TransCDR and other compared models. We reported the mean and standard deviation of metrics obtained by executing 10-CV for each method.

## 2.5 GSEA

Performing GSEA on the omics data of patients from TCGA can offer valuable biological insight into TransCDR. We employed the trained TransCDR classification model to evaluate 225 drugs on 7,675 patients with the available 3 omics profiles from TCGA. For each drug, patients were ranked based on their prediction score, and the top and bottom 5% (384 of each) were classified as drug-sensitive and drug-resistant, respectively (Figure 3). The difference in predicted score between drug-sensitive and drug-resistant patients was calculated using the formula:

$$Diff_{\text{drug}} = \bar{S}_{\text{sen}} - \bar{S}_{\text{res}} \qquad (16)$$

Furthermore, we sorted $Diff_{\text{drug}}$ in descending order to further identify the top 10 drugs to analyze the biological mechanisms underlying drug sensitivity/resistance. We calculated the log2 fold change of genes for each drug between drug-sensitive and resistant patients:

$$\log_2 FC = \log_2(\bar{X}_{\text{sen}}/\bar{X}_{\text{res}}) \qquad (17)$$

We conducted GSEA on the differentially expressed genes with $\log_2 FC$ using the *clusterProfiler* R package and Molecular Signature Database v2023, which contains 33,591 gene sets across 9 major collections. Gene sets were significantly enriched if they had both Benjamini-Hochberg corrected *p*-value and FDR *q*-value <0.01 and |*NES*|≥1.9.

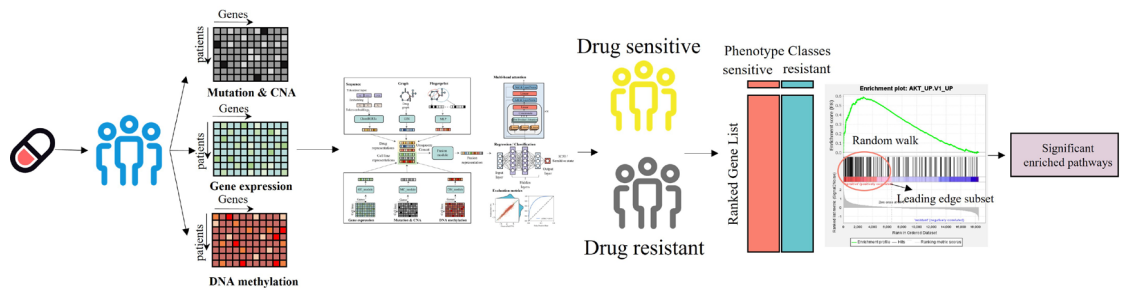

Figure 3. The workflow of GSEA. The patients are classified into drug-sensitive or drug-resistant groups using the TransCDR classification model. The GSEA method is employed to identify pathways that are significantly enriched.

## 2.6 Implementation details

All models based on neural networks, including TransCDR, DeepTTA, GraphDRP, TGSA, DRPreter, and TransCDR varieties, were developed using Pytorch. The training process was limited to 100 epochs for all training sets and models. The Adam optimizer with a learning rate of $10^{-5}$ was used to update the model parameters during the back-propagation process. The batch size was 64, MSELoss was employed as the loss function for regression models, and BCELoss for classification models. A dropout rate of 0.1 was specified, and the validation set was used to fine-tune hyperparameters and stop the training process. All experiments were conducted on Tesla A100 GPUs with 40 GB of memory. GSEA was conducted on RStudio. For further details, please refer to the respective GitHub repository: https://github.com/XiaoqiongXia/TransCDR.

# 3 Results

## 3.1 Performance evaluation of TransCDR under 5 sample scenarios

The evaluation performance of TransCDR exhibited significant variations across 5 distinct sample scenarios (i.e., warm start, cold cell (10 clusters), cold drug, cold scaffold, cold cell & scaffold), underscoring the diverse efficacy of TransCDR and its applicability in real-world contexts. For the warm start scenario, TransCDR exhibited relatively high prediction performance with an *RMSE* of 0.9703±0.0102 and *PC* of 0.9362±0.0014 in regression tasks, indicating its precise application in predicting missing $IC_{50}$ of drugs on cell lines in GDSC. However, the cold start scenario was more challenging due to the inclusion of scaffold/cell lines that were unseen during the training process. TransCDR performed worse with more strict data segmentation strategies (Figure S1). As demonstrated in Table 1, the regression *PC* of TransCDR was 0.8639±0.0103 under the strictest cold cell scenario, highlighting its generalizability in predicting drug responses of unseen omics profiles, particularly for patients with known anticancer drugs, which can greatly aid precision medicine. The *PC* values were found to be 0.5467±0.1586, 0.4816±0.1433 and 0.4146±0.1825 for cold drug, cold scaffold and cold cell & scaffold scenarios, respectively, suggesting its potential in predicting massive unseen drug/compound responses on seen/unseen cell lines, hence, offering a powerful tool for drug repurposing and discovery.

Table 1. Evaluation performance of TransCDR under the 5 scenarios.

| Sample scenarios | *RMSE* | *PC* | *SC* | *C-index* |
| --- | --- | --- | --- | --- |
| Warm start | **0.9703±0.0102** | **0.9362±0.0014** | **0.9146±0.0020** | **0.8797±0.0013** |
| Cold cell (10 clusters) | 1.3949±0.0897 | 0.8639±0.0103 | 0.8243±0.0085 | 0.8213±0.0051 |
| Cold drug | 2.2756±0.3785 | 0.5467±0.1586 | 0.4678±0.1367 | 0.6651±0.0523 |
| Cold scaffold | 2.3722±0.3794 | 0.4816±0.1433 | 0.4470±0.1423 | 0.6571±0.0522 |
| Cold cell & scaffold | 2.4518±0.4201 | 0.4146±0.1825 | 0.3681±0.1918 | 0.6283±0.0693 |

The performance of the TransCDR regression model is assessed using metrics such as *RMSE*, *PC*, *SC*, and *C-index*. All results are obtained by 10-CV.

## 3.2 Performance comparison of TransCDR and other models

To verify the effectiveness of our proposed TransCDR, we compared TransCDR with DeepCDR [8], GraphDRP [10], DeepTTA [9], TGSA [17], and DRPreter [18] on the GDSC dataset. TransCDR achieved the best performance with the highest *PC*, *SC*, and *C-index* compared to DeepCDR, GraphDRP_GAT_GCN, GraphDRP_GINConvNet, GraphDRP_GATNet, and GraphDRP_GCNNet under all scenarios (Figure 4, Figure S2). DRPreter and TGSA achieved comparable performance with TransCDR on a warm start but performed poorly under the cold scaffold and drug. The results indicated that DRPreter and TGSA were overfitting to training sets and thus cannot generalize to the novel drugs and scaffolds. TransCDR displayed superior generalization capabilities, particularly in the challenging cold scaffold task. TransCDR had comparable performance with DRPreter, TGSA, and DeepTTA under cold cell cluster, even though TransCDR was trained without prior knowledge: protein-protein interactions (Figure S2 G-I). These findings suggested that the transfer learning strategy could effectively transfer the knowledge learned from a large-scale chemical dataset, thereby improving the prediction performance of TransCDR on novel drugs and scaffolds. From the perspective of real application scenarios, TransCDR was the best model to efficiently integrate information and extract features from the structures of drugs and multi-omics data of cell lines for drug response predictions.

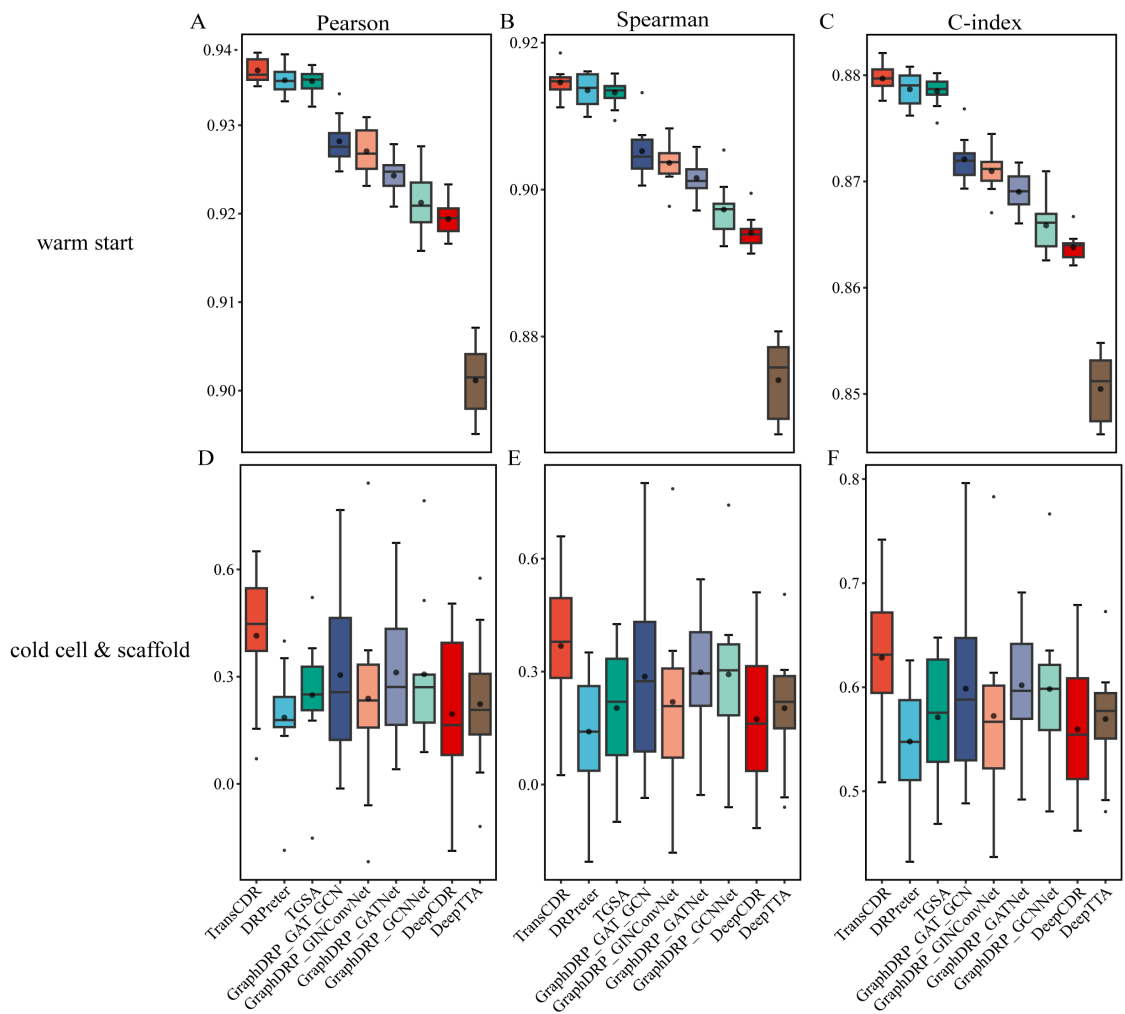

**Figure 4.** Performance comparisons are conducted between TransCDR and 8 other approaches, namely, DRPreter, TGSA, GraphDRP_GAT_GCN, GraphDRP_GINConvNet, GraphDRP_GATNet, GraphDRP_GCNNet, DeepCDR, and DeepTTA on warm start and the strictest scenario: cold cell & scaffold.

## 3.3 Transfer learning exhibits superior performance in comparison to training a model from scratch

We investigated the effectiveness of transfer learning by converting the pre-trained drug representation modules into drug encoders trained from scratch (Section 2.3.1.4). As depicted in Figure 5, TransCDR with pre-trained drug encoders demonstrated superior performance compared to its variants, including sequence-based (i.e., TransCDR_CNN and TransCDR_RNN), graph-based (i.e., TransCDR_AttentiveFP, and TransCDR_NeuralFP), and FP-based (i.e., TransCDR_ECFP) models. Specifically, the *RMSE* of TransCDR variants increased to over 0.9845, while the *PC*, *SC*, and *C-index* of TransCDR variants dropped below 0.9342, 0.9124, and 0.8780, respectively (Wilcoxon test, $p<0.05$). In addition, we examined the transfer learning performance of

TransCDR under cell line, cold drug, cold scaffold, and cold cell & scaffold scenarios, as demonstrated in Figure S3. TransCDR achieved satisfactory results as expected, outperforming TransCDR_RNN, TransCDR_CNN, TransCDR_AttentiveFP, and TransCDR_NeuralFP. Specifically, TransCDR showed a *PC* of 0.5467, superior to the second-best model TransCDR_ECFP, with a *PC* of 0.4624 under the cold drug scenario. The results suggested that transfer learning was reliable for learning drug representations by leveraging the chemical knowledge extracted from large-scale datasets like ZINC and PubChem. Notably, TransCDR variants inherently learned drug representations by training an end-to-end model on the training set. However, their performance on the test set with unseen drugs could be better. TransCDR_ECFP attained better performance than other variants thanks to the generation of informative FP representations through Morgan's algorithm.

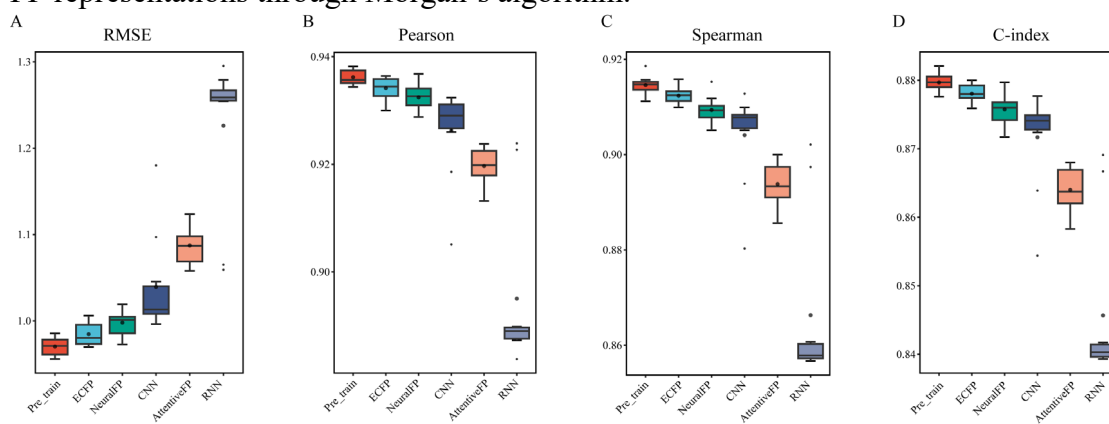

**Figure 5.** The performances of TransCDR and its variants with various drug representation modules are evaluated in a warm start scenario.

## 3.4 Impact of each modality in TransCDR

The present study provided insights into the effectiveness of the proposed framework for drug response prediction. Ablation studies were conducted by removing each feature (i.e., S, G, and FP of drugs, and MC, GE, and DM of cell lines) from TransCDR, and the resulting decrease in predictive performance was analyzed. Figure 6 demonstrated that removing these features affected the performance of TransCDR. MC was found to be the most critical among cell line features, followed by GE and DM. FP was identified as the most significant for drug features, followed by G and S. These findings corroborated the comparison results presented in Figure 6 and highlighted how multi-modality fusion could enhance model performance by complementing the limitations of individual modalities. In summary, MC and FP contributed the most among different omics profiles for cell lines and drug notations for drugs, respectively.

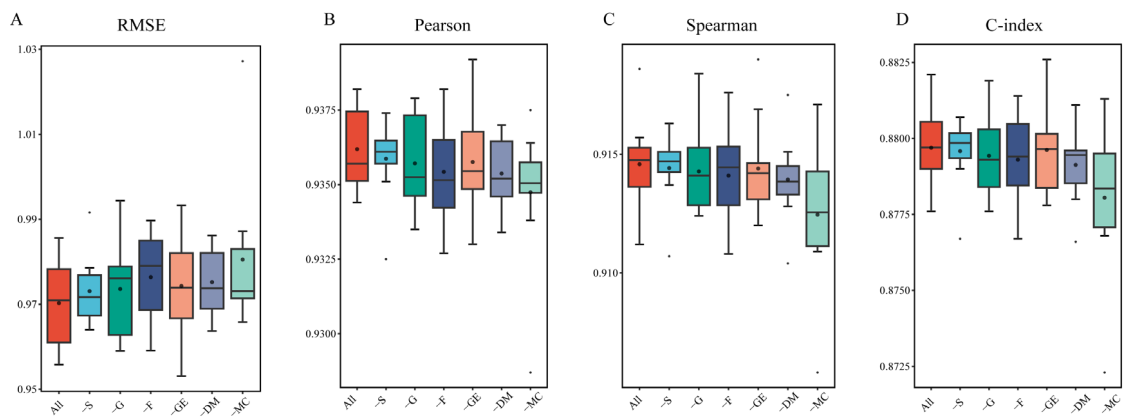

**Figure 6.** Model ablation experiments results. The x-axis denotes the removal of a specific modality. MC is the most critical characteristic of cell lines, as the exclusion of this feature significantly ($p<0.05$) increases the *RMSE* values of TransCDR compared to models without GE or DM. Similarly, FP is the most significant feature for drugs. TransCDR model without FP demonstrates significantly ($p<0.05$) increased *RMSE* values compared to that without G or S.

## 3.5 The effectiveness of self-attention

Two variants of cross-attention, DCA and CDA, were implemented along with concatenation operation to assess the efficacy of self-attention in the fusion module. Notably, self-attention outperformed other fusion methods in all regression evaluation metrics, including *RMSE*, *PC*, *SC* and *C-index* (Table 2). For instance, the *RMSE* achieved by self-attention was recorded as 0.9703±0.0102, surpassing the second-best option of concatenation, 0.9845±0.0147. On the other hand, the performance of DCA and CDA was inferior as they only focused on the cross-effect between drugs and cell lines while disregarding the internal feature interaction of either. Thus, the self-attention-based fusion module was employed to fuse multi-modal features.

Table 2. The performance of TransCDR with distinct fusion methods.

| Fusion module | *RMSE* | *PC* | *SC* | *C-index* |
| --- | --- | --- | --- | --- |
| Self-attention | **0.9703±0.0102** | **0.9362±0.0014** | **0.9146±0.0020** | **0.8797±0.0013** |
| DCA | 0.9962±0.0208 | 0.9326±0.0029 | 0.9099±0.0025 | 0.8761±0.0021 |
| CDA | 1.0303±0.0104 | 0.9275±0.0015 | 0.9048±0.0024 | 0.8720±0.0017 |
| Concatenate | 0.9845±0.0147 | 0.9339±0.0020 | 0.9117±0.0022 | 0.8773±0.0017 |

The best results are emphasized using bold font, and the second-best results is underlined.

## 3.6 TransCDR predicts binary drug response

We subsequently assessed the predictive power of TransCDR in cell line responses to drugs. TransCDR demonstrated high performance across varying ratios of positive and negative samples in the warm start scenario. Specifically, when the dataset was balanced, TransCDR yielded superior performance with an *AUROC* of 0.8213±0.0067 and an *AUPR* of 0.8138±0.0085. When the dataset was unbalanced of 1:2, 1:5, and 1:8,

TransCDR displayed a slight increase in *AUROC* and a decline in *AUPR*, with reductions over 8.76%, 20.99% and 26.93% for *AUPR*, respectively. These findings highlighted the impact of dataset imbalance on the predictive power of TransCDR, with *AUPR* exhibiting sensitivity to sample ratio variations. Therefore, we utilized *AUPR* as the primary evaluation metric. In the cold test setting, the *AUPR* of TransCDR reduced more compared with a warm start when the dataset was imbalanced. Specifically, when the sample ratio was 1:1, TransCDR in cold cell achieved an *AUPR* of 0.7492±0.0227, which was 39.52% higher than that of the 1:8 sample ratio of (*AUPR*=0.3540±0.0381). Similarly, in the cold drug setting, a sample ratio 1:1 yielded optimal performances (Figure 7). Consequently, subsequent experiments were conducted using the sample ratio of 1:1.

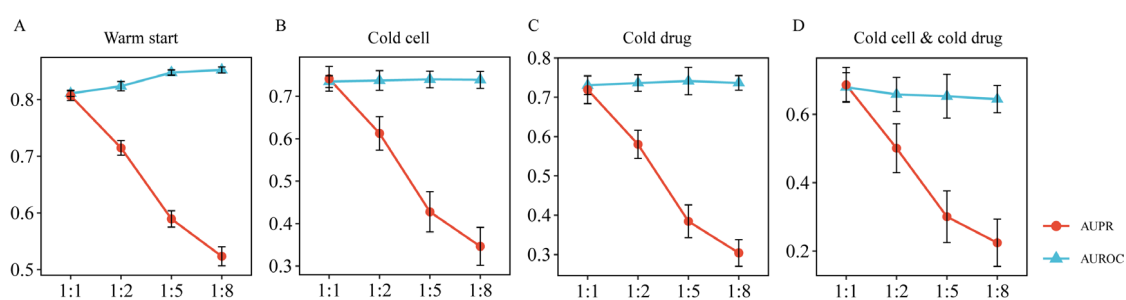

**Figure 7.** TransCDR's performance evaluation is assessed across 4 sampling scenarios utilizing 4 sampling ratios between positive and negative samples (1:1, 1:2, 1:5, and 1:8).

## 3.7 Application of TransCDR on GDCS

The pre-trained TransCDR exhibited excellent performance across a diverse range of cancer types (Figure 8A-C), cell lines (Figure 8D-F), and drugs (Figure 8G-I). In all tested cancer types, the *PC* and *SC* values ranged from 0.9624 to 0.9763 and 0.9345 to 0.9591, respectively (Table S1). The *PC* and *SC* values for cell lines ranged from 0.9192 to 0.9886 and 0.8723 to 0.9676, respectively (Table S2). The performance of TransCDR on drugs varied considerably, with the *PC* and *SC* ranging from 0.3949 to 0.9838 and 0.3983 to 0.9814, respectively (Table S3).

Employing the trained TransCDR, we predicted 34,662 missing $IC_{50}$ values for drug-cell line pairs in the GDSC database, corresponding to approximately 18.10% of all 191,475 pairs involving 851 cancer cell lines and 225 drugs. We ranked the $IC_{50}$ values predicted by the regression model in ascending order and selected the top 10% (3,466) drug-cell line pairs inclusive of 610 cancer cell lines and 75 drugs (Table S4). Our work confirmed previous research findings on the top 15 (the lowest $IC_{50}$) drug-cell line pairs that were molecularly effective in cancer treatment, involving 15 cell types, 4 drugs, 10 tissues, and 8 cancers (Table S5). Notably, Bortezomib, one of the approved proteasome inhibitors for treating various malignancies (e.g., SKCM, OV and BRCA) [36], was predicted to be sensitive to different cell lines and cancer types. The top 10 'sensitive' and the last 10 'resistant' drugs were depicted in Figure 8J. As anticipated, several sensitive/resistant drugs were also identified by DeepCDR [8]. For instance,

Bortezomib, docetaxel, epothilone B, vinblastine, vinorelbine, and SN-38 [37] were predicted as sensitive drugs, and FR-180204, NSC-87877, GW-2580, DMOG, phenformin, and AICAR were predicted as resistant drugs by DeepCDR. Additionally, the effectiveness of the most potent drugs, Bortezomib [38], docetaxel [39], and vinblastine [40], has been established in multiple cancer types.

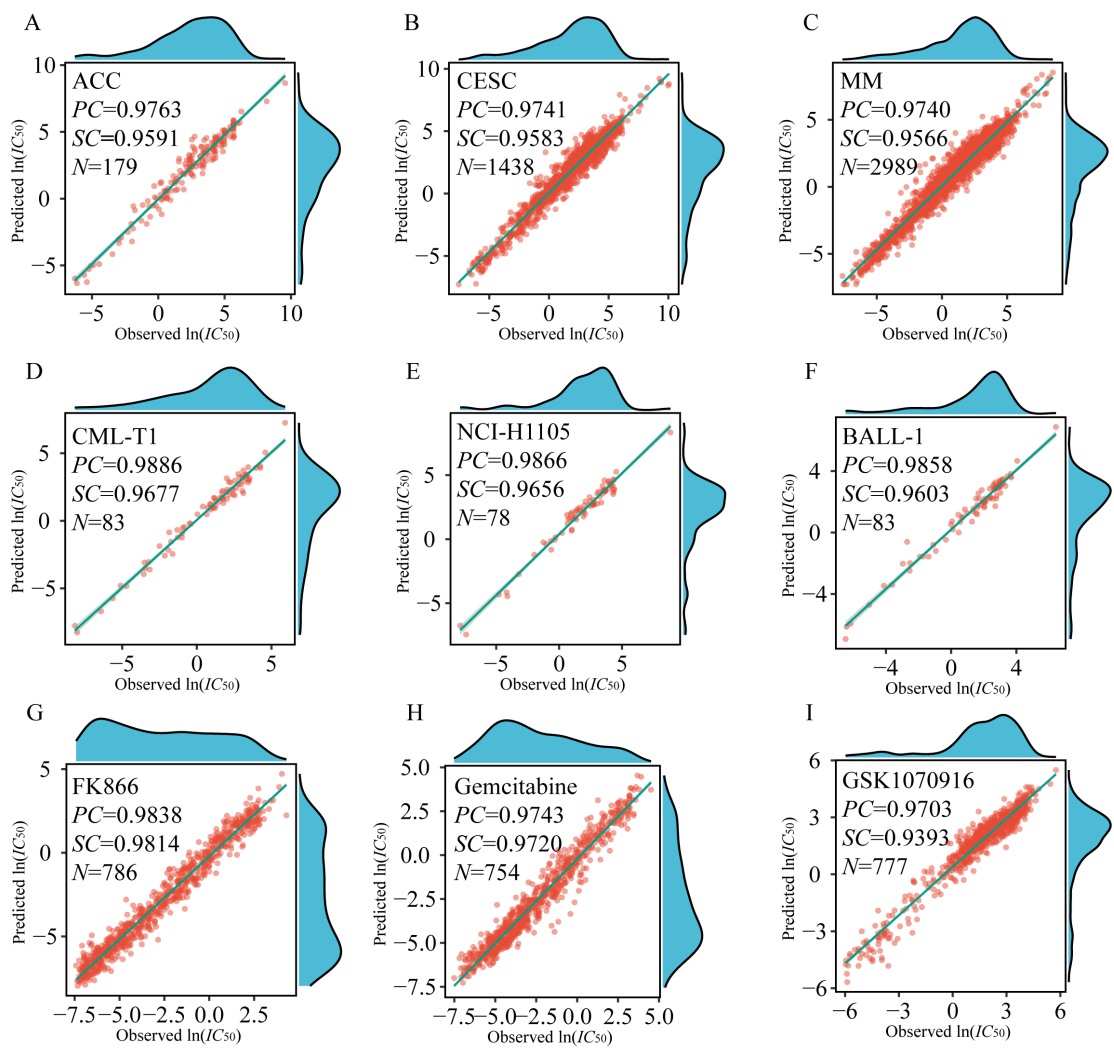

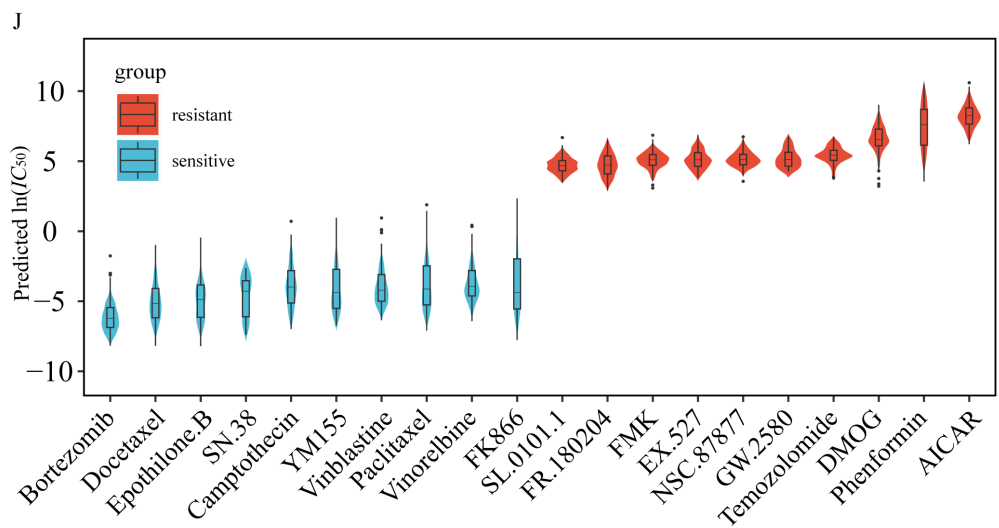

**Figure 8.** (A-C) The scatter plots of CDRs of specific cancer types, specifically adrenocortical carcinoma (ACC), cervical squamous cell carcinoma and endocervical adenocarcinoma (CESC), and multiple myeloma (MM), with the top 3 prediction performances. (D-F) The scatter plots of CDRs of specific cell line types, specifically T-lymphoid cell line (CML-T1), NCI-H1105, and BALL-1, with the top 3 prediction performances. (G-I) The scatter plots of CDRs of specific drugs, including FK866, Gemcitabine, and GSK1070916, with the top 3 prediction performances. (J) The study categorizes drugs based on their average predicted $IC_{50}$ values in ascending order, with the top 10 drugs being sensitive and the bottom 10 being resistant.

## 3.8 External validation results

The present study assessed the efficacy of TransCDR trained on the GDSC dataset by evaluating the external in vitro dataset CCLE. The results demonstrated TransCDR's outstanding performance with a *PC* range varying from 0.6279 to 0.8926 when tested across diverse cancer types. Bile duct cancers exhibited the highest performance (*PC* of 0.8926), while kidney cancer demonstrated the lowest (*PC* of 0.6279) (Table S6). These findings suggested that TransCDR could effectively predict drug response in new cell lines specific to certain cancer types.

## 3.9 TransCDR recognizes biological mechanisms under drug response

To recognize biological mechanisms under drug response, we utilized the pre-trained TransCDR to screen 225 drugs for 7,675 patients from TCGA. The predicted drug sensitivities of these patients were presented in Table S7. We selected the top 10 drugs, namely CX-5461, Lapatinib, Dasatinib, Erlotinib, Afatinib, Trametinib, Nutlin-3a, A-770041, CHIR-99021, and AZD-0530 for GSEA. By performing GSEA, we were able to elucidate the biological mechanisms underlying the predicted drug sensitivities of patients and explore possible underlying mechanisms (Table S8). Our observation showed that differential expression genes caused by Afatinib medication demonstrated a significant enrichment within gene sets associated with breast and lung cancer. This observation aligned with evidence supporting Afatinib's efficacy in treating breast and lung cancer [41, 42]. Furthermore, the enriched gene sets offered insight into Afatinib's therapeutic mechanisms. For instance, up-regulated genes observed in Afatinib-sensitive patients exhibited a significant enrichment in COLDREN_GEFITINIB_RESISTANCE_DN (*NES*=1.983, *p*=0.0005), which pertained to genes that down-regulated in non-small cell lung carcinoma cell lines resistant to Gefitinib in comparison to those that were sensitive [43]. This finding indicated that Gefitinib and Afatinib operated through similar mechanisms [44]. In contrast, the up-regulated genes observed in Afatinib-sensitive patients showcased a significant enrichment in HOLLERN_EMT_BREAST_TUMOR_DN (*NES*=2.211, *p*=4.53E-06), which consisted of genes with low expression levels in mammary tumors marked by epithelial-mesenchymal transition histology and could result in resistance to Afatinib [45].

# 4 Discussion

In comparison with existing SOTA models, TransCDR exhibited several improvements. Firstly, it outperformed other models across diverse prediction tasks under different sample scenarios (warm and cold start). Secondly, TransCDR fused the most extensive data modalities, incorporating 3 drug representations and 3 omics profiles, whereas DeepTTA only considered SMILE strings and gene expression profiles. Thirdly, TransCDR learned the fusion representations by a self-attention-based module which was more effective than a simple concatenation operation. Thirdly, we comprehensively assessed the generalizability of TransCDR across diverse scenarios. Our proposed model enhanced the performance in cold drug/scaffold and cold cell & scaffold scenarios, essential for predicting cancer drug response and screening novel candidates from a vast drug/compound space.

We demonstrated that generalizing TransCDR to novel scaffolds posed a greater challenge than cell line clusters. Several factors contributed to this phenomenon. Cell lines were characterized by gene expression profiles obtained via omics measurements, providing a comprehensive representation of cellular biology features. Conversely, compounds were encoded using SMILES strings, which may lead to loss of structural information. Furthermore, TransCDR learned drug embedding from SMILES strings or molecular graphs using end-to-end training, requiring substantial drug structures. Lastly, minor structural differences between similar compounds may result in significant disparities in SMILES strings, yielding distinct embeddings. TransCDR can serve as an effective tool for the cancer-drug response prediction. Additionally, TransCDR have promising applications in drug discovery. Specifically, we can initially assess the scaffold similarity of a new compound/drug against known drugs; if a similarity scaffold is identified, our predicted CDRs will hold greater credibility. If not, TransCDR stands as the optimal model to predict CDRs in cold scaffold and cold cell & scaffold scenarios.

However, several limitations and potential directions for further improving TransCDR have been identified. The study requires large-scale, highly qualified datasets, including multiple drugs and cell lines. Although drug response data have increased dramatically over the past decades, cell lines with multi-omics profiles are limited. The performance of TransCDR on the cold scaffold is significantly better than other SOTA models through transfer learning but still has much room for improvement. The current TransCDR cannot capture the drugs' three-dimensional structural information, which inevitably affects drug representation learning. A better drug representation model that can extract discriminating features from drug notations will be designed, such as GeoGNN, which encodes molecules' topology and geometry information by a geometry-based GNN architecture [54]. Therefore, to further improve the prediction performance and interpretability of TransCDR, we will propose the next version of TransCDR, trained on the larger and more reliable CDR dataset, considering the multimodal features of drugs and cell lines and making full use of prior domain knowledge.

# 5 Conclusions

In this study, we presented an end-to-end CDR prediction model called TransCDR, which fused multi-modality representations of drugs, including SMILES string, molecular graph, ECFP, and omics profiles, including genetic mutation, gene expression, and DNA methylation to learn the $\ln(IC_{50})$ values or sensitive states of drugs on cell lines. TransCDR outperformed the 8 SOTA models and showed high performance under different sample scenarios. In addition, TransCDR outperformed multiple variants with drug encoders that were trained from scratch. We confirmed that FP and genetic mutation contributed the most among multiple drug notations and omics profiles, respectively. Furthermore, TransCDR showed high prediction performance on the external test sets CCLE. Finally, we predicted $\ln(IC_{50})$ values of missing CDRs in GDCS and screened the drug response of cancer patients to drugs. These candidate CDRs were verified by existing literature and GSEA. In summary, our deep learning model, TransCDR, offers a powerful tool for drug response prediction.

# Funding

This work was supported by the Peak Disciplines (Type IV) of Institutions of Higher Learning in Shanghai.

# Declaration of Competing Interest

The authors have declared that no competing interests exist.

# Data Availability Statement

The supplementary appendixes, tables and figures that mentioned in this paper can be found online. The Python and Torch implementation of the TransCDR model is accessible at https://github.com/XiaoqiongXia/TransCDR.

# Acknowledgements

We acknowledge the support provided by the project of Shanghai's Double First-Class University Construction, the Development of High- Level Local Universities: Intelligent Medicine Emerging Interdisciplinary Cultivation Project, and Medical Science Data Center of Fudan University.